# Regional poverty in Bulgaria in the period 2008-2019

Iva Raycheva[1]

[1]*Department of Statistics and Econometrics, Faculty of Applied Informatics and Statistics,*
*University of National and World Economy, 1700 Sofia, Bulgaria*

**Abstract:**
**Background**: Poverty among the population of a country is one of the most disputable topics in social studies. Many researchers devote their work to identifying the factors that influence it most. Bulgaria is one of the EU member states with the highest poverty levels. Regional facets of social exclusion and risks of poverty among the population are a key priority of the National Development Strategy for the third decade of 21st century. In order to mitigate the regional poverty levels it is necessary for the social policy makers to pay more attention to the various factors expected to influence these levels.
**Materials and Methods:** In the present study, in addition to gaining insights from theories, a statistical method known as "stepwise regression" is used to select factors that best describe the variation of regional poverty levels. Panel data analysis methods (fixed and random effects models) are applied as well. For this purpose, a range of variables for Bulgarian districts and studied period are extracted for generating the necessary panel data basis. Two regression models are estimated using this data: the first one reflects the employment level in the districts as a potential factor influencing regional poverty; the second one implements the stepwise regression analysis by which three indicators that significantly affect poverty levels were identified: "Expenditure on the acquisition of tangible fixed assets", "Employment", and "Housing stock".
**Results**: Poverty reduction is observed in most areas of the country. The regions with obviously favorable developments are Sofia district, Pernik, Pleven, Lovech, Gabrovo, Veliko Tarnovo, Silistra, Shumen, Stara Zagora, Smolyan, Kyustendil and others. Increased levels of poverty are found for Razgrad and Montana districts. It was fond that the reduction in the risk of poverty is associated to the increase in employment, investment, and housing.
**Conclusion:** The social policy making needs to be aware of the fact that the degree of exposition to risk of poverty and social exclusion significantly relates to the levels of regional employment, investment and housing.
**Key Words**: Regional poverty; Employment; Panel data analysis; Bulgaria.

## I. INTRODUCTION

One of the most discussed indicators utilized for the goals of social policy making is the level of poverty among the population. Many researchers dedicate their studies to measuring, describing, and analysing poverty domestically and around the world. Poverty is officially defined as „*a condition characterized by severe deprivation of basic human needs, including food, safe drinking water, sanitation facilities, health, shelter, education and information. It depends not only on income but also on access to services*" (UN, 1995). Palmer, Rahman, and Kenway summarized a general perception of poverty that it: „*is concerned with a lack of possessions, or ability to do things that are, in some sense, considered 'normal' or 'essential' in society*" (Palmer et al., 2002).

Identifying the factors influencing poverty is one of the most disputable issues, in particular, the appropriate ways to tackle this multifaceted problem. It is generally believed that „*cash income is a key factor, but is not the only indicator of people's access to goods and services*" (Gordon et al., 2000). For example, the fields of social exclusion are defined by De Haan and Maxwell (1998) as three key arenas and their intrinsic elements: (a) rights: human, legal/civic, democratic; (b) resources: human and social capital, labor markets, product markets, state provision, common property resources; (c) relationships: family networks, wider support networks, voluntary organisations.

Among the most frequently studied factors determining poverty are the remuneration and employment levels. MacInnes et al. (2013) conducted a study in which they look a little deeper into the problem – these authors outline that „*changes to the welfare system have made poverty worse. The small, positive developments in the labor market cannot in any way be seen as 'balancing' reforms that have cut the incomes of some of the poorest people in the country*". According to a study by Tinson et al. (2016), in addition to income, housing is another major factor influencing poverty in England. They raise questions about the prices and quality of housing, and also pay attention to whether people live in rented accommodation or have their own housing. These authors emphasize that „*housing in the UK is too often expensive and of poor quality, particularly in the private rented sector. Work, although the best defence against low income, is too often insufficient. The social security system has become less effective for those with housing costs and Council Tax to pay*" (Tinson et al., 2016).

Bulgaria is one of the countries in the European Union with the highest levels of poverty. In his study Ismailov (2019) points out some of the factors that have influenced the poverty level in the country, e.g. „*the worsened social indicators such as educational level, health status, access to basic goods and services, and index of human development*". Many studies of the social policy in Bulgaria have been focused on the approaches to alleviation of poverty. Among the priorities of the National Strategy towards the reduction of poverty is the provision of enhanced employment opportunities and increases remuneration level "*through active involvement of citizens in the labor market*", in particular, by encouraging entrepreneurship (Terziev et al., 2016). This way, the major levers which social policy relies on are the minimum wage and employment policies. However, despite the efforts of the institutions the poverty levels in the country still stay among the highest in the European Union. It follows that either the measures targeted to these two factors are not effective, or additional factors that affect the poverty level in the country must be sought – especially, in the most affected regions of the country.

Mintchev et al. (2010) indicate as one of the reasons for increased levels of poverty in Bulgaria the persisting poverty among Roma families, especially in respect of the multi-child Roma households with chronical unemployment. Empirical evidence is provided concerning the divergence of the effects from social assistance measures. In her study Zhelyazkova (2018) also tries to characterize poverty and the reasons for having an improper approach to its alleviation in Bulgaria. Albeit the country experiences economic growth and substantial relative increment of the average wage during the second decade of 21st century, "*the deformed system of social assistance leads to an increase in income distances and therefore leads to an even greater depth of poverty*" (Zhelyazkova, 2018). Stoyanova, Kokova, and Stoev (2020) indicated the unemployment as a traditional factor of poverty influencing mostly groups at-risk as: "*young people with or without disabilities; persons... who are illiterate or with primary or lower education, without professional qualification, or lack of key competencies; economically inactive persons searching for jobs under any conditions (forming the group of the so-called working poor); persons living in poor housing conditions*" (Stoyanova et al., 2020). In particular, substantially affected are single parent families that have limited resources and experience difficulties in meeting their basic needs, as well as multi-child families.

## II. MATERIAL AND METHODS

The data used in the current study is provided basically from the information system of the National Statistical Institute of Bulgaria. The utilized statistical data is annual, reflecting the demographic and socio-economic conditions of the administrative regions (districts) at NUTS-3 level. Some of the variables are extracted directly from the database platform and others are transformed for the aim of the current research. The data refers to all 28 Bulgarian districts for the period 2008-2019.

1. The variable „People at risk of poverty or social exclusion" is an indicator that captures the level of poverty in the district. This is the main indicator used by Eurostat in poverty analyzes. The current study utilizes data on this indicator for all 28 districts for every year from the period 2008-2019.
2. Average annual wages and salaries of the employees under labor contract (with acronym AAWage)
3. Expenditure on acquisition of tangible fixed assets, in Thousand BGN (TFA)
4. Employed, in Thousand persons (Empl)
5. Unemployed, in Thousand persons (Unempl)
6. Residential buildings, in Number (RB)
7. Dwellings, in Number (D)
8. Accommodation establishments, in Number (AE)
9. GDP per capita, in BGN (GDPpc)
10. Dwellings with one room, in Number (DNR1)
11. Medical personnel: Physicians, in Number (P)
12. Museum visits, in Number (MV)
13. Population, in Number (N)
14. People convicted in commitment of crime, total, in Number (PTC)
15. People convicted by place of commitment of crime, three and more crimes, in Number (P3C)
16. Crimes with penalty inflicted, in Number (C)
17. National road network, Category Motorways, in km (RM)
18. National road network, Category I Roads, in km (R1)

19. Students in general and special schools, in Number (S)
20. Teachers in general and special schools, in Number (T)

This study examines the relationship between the regional poverty rate and the factors expected to affect it in the period 2008-2019 in Bulgaria. The stepwise regression method is used to select the independent variables that correlate significantly with the poverty level in the 28 Bulgarian districts for the period. This method utilizes the power of iterative algorithms to sequentially introduce variables in the multiple regression model in order to derive an optimal group of regressors, regardless of the sequence of their inclusion (Boshnakov et al., 2020). Additionally, panel regression models are estimated with fixed effect and random effect specifications (Wooldridge, 2013). To take into account the panel nature of the data, the fixed effects model introduces two groups of dummy variables: district ("fixed" for each region, regardless of time) and time dummies ("fixed" for each year, regardless of region).

## III. RESULT

*Comparative analysis: „People at risk of poverty or social exclusion"*

In this study, the variable „People at risk of poverty or social exclusion" was initially reviewed through the prism of static data (looking at all 28 regions of Bulgaria) for the year of 2008 firstly, and then for year 2019. Figure 1 presents the indicator for 2008 by Bulgarian districts (percentages indicate the share of people at-risk-of poverty or social exclusion). The map shows that the highest percentages of the levels of people at risk of poverty or social exclusion are in the districts of Yambol with 57%, Sliven with 53% and Veliko Tarnovo with 50.9%. The districts with the lowest values of the percentage of people at risk of poverty are Blagoevgrad with 23.9%, Sofia capital with 28.6% and Plovdiv with 30.9%.

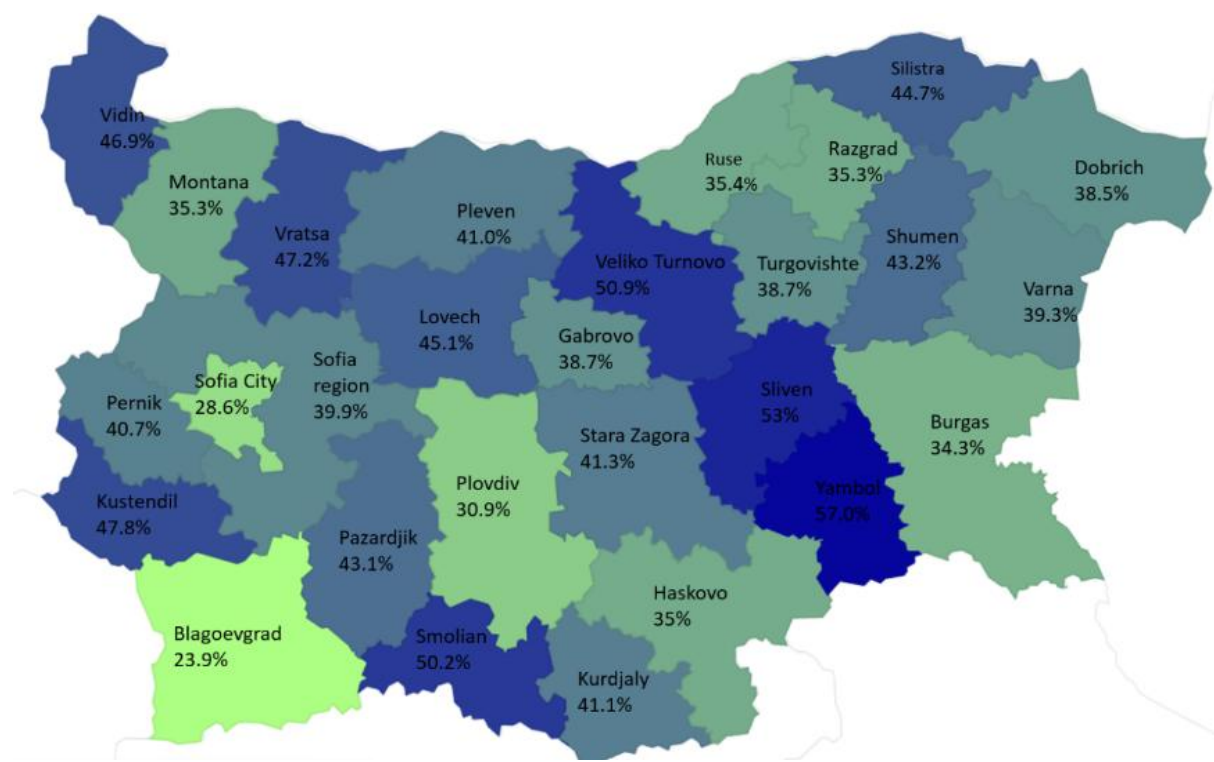

**Figure 1.** „People at risk of poverty or social exclusion" for 2008.

Figure 2 shows the variable „People at risk of poverty or social exclusion" for the 28 districts in 2019. The map shows that for 2019 the highest shares of people at risk of poverty or social exclusion are in the districts of Razgrad with 48.8%, Yambol with 47.6% and Sliven with 43.5%. Yambol and Sliven districts continue to maintain very high levels of poverty. The districts with the lowest values of the percentage of people at risk of poverty are Sofia district with 25%, Sofia capital city with 26%, and Gabrovo with 26.8%. The capital city of Sofia continues to be one of the districts with the lowest levels of poverty in the country.

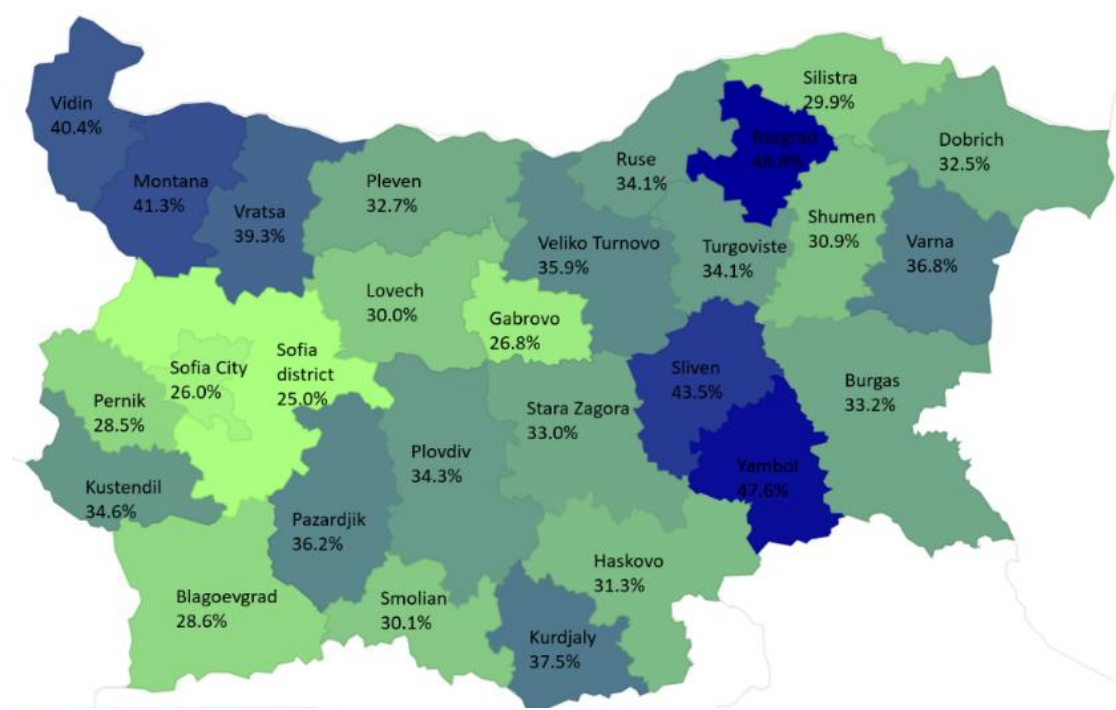

**Figure 2.** „People at risk of poverty or social exclusion" for 2019.

Figure 1 and Figure 2 clearly show the change in the color and percentage of the poverty indicator in some of the areas. Significant tackling of poverty is seen in some areas. For example, Sofia district in 2008 had a value of 40% and in 2019 it dropped to 25% - this way, relocated from a district with average poverty level to a district with one of the lowest levels of the indicator. There are other positive examples of a significant reduction in the percentage values of the variable, such as:

- ❖ Veliko Tarnovo - from 51% decreases to 36% - an example for a districts among those with highest levels of poverty that improves to a district with medium level of poverty;
- ❖ Smolyan - from 50.2% decreases to 30.1%;
- ❖ Kyustendil - from 47.8% decreases to 34.6%;
- ❖ Lovech - from 45.1% decreases to 30.0%;
- ❖ Silistra - from 44.7% decreases to 29.9%;
- ❖ Shumen - from 43.2% decreases to 30.9%;
- ❖ Stara Zagora - from 41.3% decreases to 33.0%;
- ❖ Pleven - from 41.0% decreases to 32.7%;
- ❖ Pernik - from 40.7% decreases to 28.5%;
- ❖ Gabrovo - from 38.7% decreases to 26.8%.

However, in other districts the reverse process of increasing poverty is observed. Such examples are:

- ➢ Razgrad - from 35.3% increases to 48.8%; in 2019 it is already the district with the highest share of people at risk of poverty or social exclusion;
- ➢ Montana - from 35.2% increases to 41.3%.

There are differences in the level of poverty also in the other districts, but they are negligible and would not significantly change the poverty status of the region. In general, it can be argued that poverty levels are declining in most areas during the last decade. However, the descriptive review identified two districts with a special shifts in terms of poverty – these are Razgrad and Veliko Tarnovo. They are selected for a descriptive review of the dynamics of the variable to be monitored throughout the study period. Figure 3 presents data for „People at risk of poverty and social exclusion" in Veliko Tarnovo and Razgrad districts for the period 2008-2019.

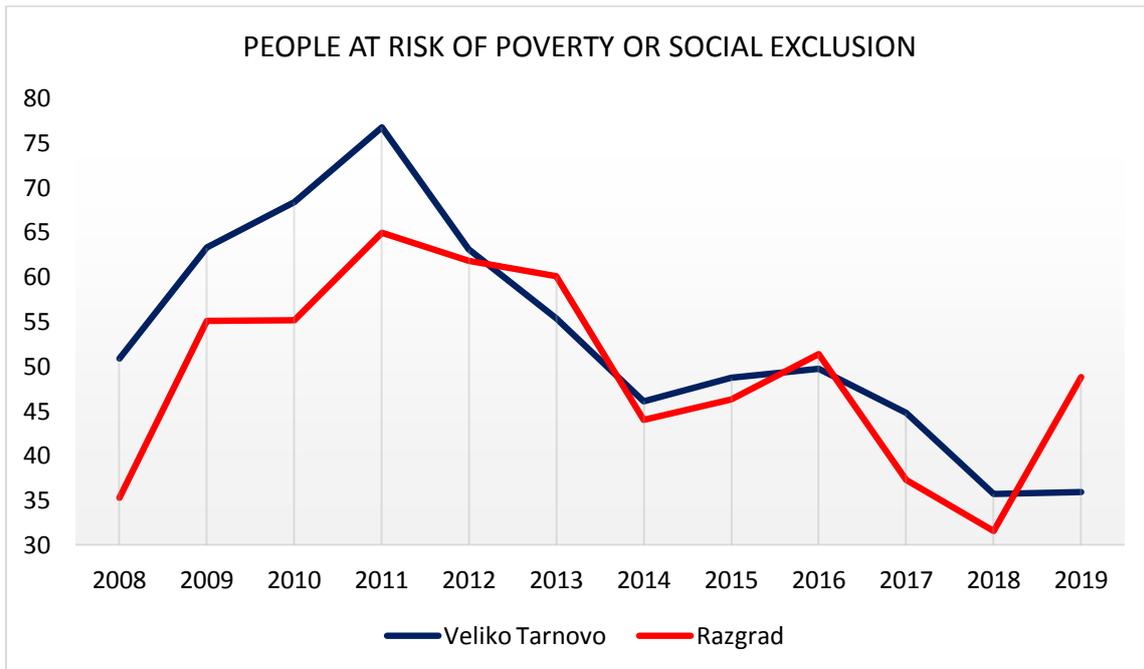

**Figure 3.** People at risk of poverty or social exclusion in Veliko Tarnovo and Razgrad, 2008-2019.

From 2008 to 2011 there was an almost parallel growth in the values of the indicator, where in Razgrad district there was a much lower initial value as compared to Veliko Tarnovo. This growth may be due to overall effect of the Global financial crisis. From 2011 to 2019, poverty decreased in both regions which is mainly due to the process of crisis recovery and economic stabilization after the Bulgarian integration in the European Union, as well as some poverty reduction policies. During this period, the two lines intersect continuously and there is no clear opinion on which district performs better. But in 2019, the percentage of the poor in Razgrad has significantly increased above that in Veliko Tarnovo, becoming the district with the highest level of poverty in the country at the end of the decade.

*A panel data analysis of the poverty in Bulgaria*

This study examines the relationship between the regional poverty rate and the factors affecting it in the period 2008-2019 in Bulgaria. For this purpose panel data regression models are applied. Here initially a bivariate model is estimated where the dependent variable is "People at risk of poverty or social exclusion" ($PRP_{it}$) and the independent variable: Employed ($Empl_{it}$). After estimating sequentially a fixed-effect and a random-effects model, the fixed-effects model was selected as more appropriate as a result of the Hausman test. Due to presence of autocorrelation and heteroscedasticity found in the residuals, the standard errors of parameter estimates have been recalculated as robust (HAC). The specification of the "fixed effects" model has the following form:

$$PRP_{it} = \beta_0 + \beta_1 Empl_{it} + \sum_{r=2}^{28} \delta_r DR_{it} + \sum_{s=2}^{12} \gamma_s DT_{it} + \varepsilon_{it} \qquad (1)$$

**Table No 1.** Results from the bivariate model with fixed effects

| Variables | Regression coefficients | | Student's test | P-value |
|---|---|---|---|---|
| | B | SE(B) robust | | |
| Constant $\beta_0$ | 196.417 | 25.286 | 7.768 | <0.000 |
| **$\beta_1$ Empl[i,t]** | ***-0.787*** | **0.212** | **-3.708** | **0.001** |
| Dependent variable: PRP[i,t]    N.T=336    R-squared=0.51 | | | | |
| Diagnosis of panel components: | | | | |
| 1) Robust F-test for joint significance on regional dummies: $H_0: \delta_r = 0$   F = 16.6   p-value=0.000 | | | | |
| 2) Wald joint test on time dummies: $H_0: \gamma_s = 0$   Chi-square(11) = 147.6   p-value = 0.000 | | | | |

In Table 1 the results from the bivariate panel model are presented. The coefficient of determination is 0.51 (i.e. 51% of the variation in "People at risk of poverty or social exclusion" can be explained by the variation of regional employment under labor contract, along with unobserved fixed regional and time effects).

The test about the parameters of the dummy variables for the regional units (deltas) shows that the null hypothesis of insignificance of these parameters can be rejected at 1% risk of error (p-value<0.01). The test about the parameters of the time dummy variables (gammas) also shows that the null hypothesis can be rejected at 1% level of significance.

The results from Student's test for the regression coefficient β1 give reason to accept that it is statistically significant at a very low level (p-value<0.01). Its value shows that an increase in the employed with 1 thousand people is associated, on average, with a decrease in the share of "People at risk of poverty or social exclusion" by about 0.8 percentage points (after controlling for unobserved individual and time fixed effects). This indicates that larger districts – attracting population and labor force – are on average characterized by lower poverty levels.

Hereafter the results from a multivariate analysis are presented – a study in which 15 possible predictor variables were expected to have impact on the share of people at risk of poverty or social exclusion in the regions. These variables have been selected to represent different economic and social areas of indicators in order to be able to cover a wide range of socio-economic influences. Next, the stepwise regression method was implemented to select – out of them – statistically significant factors that affect the poverty level in the 28 Bulgarian district for the period 2008-2019. This way, four predictor variables were selected: "Employed"; "Expenditure on acquisition of tangible fixed assets"; "Residential buildings"; "One-room dwellings".

The selection confirmed that the number of employed is retained as one of the most important predictors influencing the district level of the "population at risk of poverty". Additionally, three more variables were selected which are not unnoticed in the poverty literature. The variable "Expenditure on acquisition of tangible fixed assets" reflects the impact of investment activity at the regional level – it shows that the amount of investment is a significant factor for decreasing the level of poverty in the districts. The other two variables – Residential buildings and Dwellings with one room – are representatives of the housing stock. It turns out that the number of residential buildings in the district as well as whether homes with only one room predominate, are among the factors that affect the poverty and social exclusion level (albeit marginally significant: at 0.10 level).

Here the fixed-effects model has been identified as more efficient than random-effects model using the Hausman test. Again, due to the presence of serial correlation and heteroscedasticity in the residuals, the standard errors have been recalculated as robust (HAC). The specification of the model has the following form:

$$PRP_{it} = \beta_0 + \beta_1 Empl_{it} + \beta_2 TFA_{it} + \beta_3 RB_{it} + \beta_4 DNR1_{it} + \sum_{r=2}^{28} \delta_r DR_{it} + \sum_{s=2}^{12} \gamma_s DT_{it} + \varepsilon_{it} \quad (2)$$

**Table No 2.** Results from multivariate model with fixed effects

| Variables | Regression coefficients | | Student's test | P-value |
|---|---|---|---|---|
|  | B | SE(B) robust |  |  |
| Constant $\beta_0$ | 13,955 | 131,203 | 0,106 | 0,916 |
| **β1 Empl[i,t]** | **-1.010** | **0.205** | **-4.930** | **<0.000** |
| **B2 TFA[i,t]** | **-0.000024** | **0.000004** | **-5.649** | **<0.000** |
| β3 RB[i,t] | *0.00108* | *0.00058* | *1.862* | *0.064* |
| β4 DNR1[i,t] | *0.00283* | *0.00167* | *1.695* | *0.091* |
| Dependent variable: PRP[i,t]   N.T=336   R-squared=0.65 | | | | |
| Diagnosis of panel components: | | | | |
| 1) Robust F-test for joint significance on regional dummies: $H_0: \delta_r = 0$   F = 14.4   p-value=0.000 | | | | |
| 2) Wald joint test on time dummies: $H_0: \gamma_s = 0$   Chi-square(11) = 332.6   p-value = 0.000 | | | | |

The coefficient of determination of the multivariate model shows obviously higher coefficient of determination (65% as compared to 51% of the bivariate model) – this improvement of the explanatory power is due to the inclusion of omitted relevant variables as a result of the stepwise procedure. The dummy variable parameter test for regional units shows that the null hypothesis of insignificance of the delta parameters can be rejected at 1% risk of error (p-value<0.01). The joint test for the gamma parameters of the time dummies also shows that there is a set of strongly significance parameters at 0.01 level as well.

Student test indicates that the parameters on "Employment" and "Expenditure on acquisition of tangible fixed assets" are strongly significant at 1% level (p-value <0.001). The other two variables – "Residential buildings" and "Dwellings with one room" – have been selected by the stepwise procedure at marginal level of significance (p-value <0.10). The multivariate estimate of the net effect of the employment variable shows even a larger value: the increase in the annual employment in the districts by one thousand employees is associated to a decrease in the level of "People at risk of poverty or social exclusion" by about 1 percentage point, on average per district, other things equal. This way, the multivariate model confirms the important role of the regional employment for poverty alleviation. The increase in the annual levels of investments in the districts by one thousand BGN is associated with a decrease of about 0.000024 percentage points. The value can be rescaled, so that 10 Mn BGN increment of the regional investment per district can lead to 0.24 pp (about one quarter of a percentage point) reduction in the poverty indicator. This result gives grounds to argue that an intensified investment activity in the regions could also act as an important factor contributing to the reduction of the regional poverty levels.

The empirical results about the two indicators for the housing stock indicate that the provision of appropriate housing also matters in respect of the poverty and social exclusion in Bulgarian districts. Moreover, during the last two decades in Bulgaria a significant expansion of construction of residential buildings has been observed. However, meeting the needs of the poor for acceptable accommodation might not be directly related to the expansion of private construction sector. Many questions arise in this respect, e.g. the majority of construction companies require advance payment from customers; substantial share of the newly released dwellings are supplied against market rent; how affordable is for the families in need to get mortgage loans, etc. A variety of such research questions are to be explored in the future.

## IV. CONCLUSION

The level of poverty in the Bulgarian regions is a social issue that requires special attention of empirical research. This article traces the regional changes in terms of poverty in Bulgaria for 2008-2019. This is a period since the EU integration of the country in 2007 and also the start of the Global financial crisis. The analysis reveals that most of the NUTS-3 level districts in Bulgaria have achieved obvious reduction in poverty levels, however, some district have deteriorated their poverty status. For the purposes of the study two regression models were estimated using panel data for these districts. The first one indicated that the employment is a significant factor with an impact on the regional poverty level. The second one utilized the results from stepwise selection of predictors in a multivariate regression model, by which four variables with significant net effects have been selected. Additionally to the employment variable, the regional investment variable has been identified as strong correlate with the regional share of people at risk of poverty. Additionally, two variables reflecting the regional housing stock indicated to have a significant but modest impact on the district level of poverty and social exclusion.

This study has opened a lot of space for many questions to be explored in the future about the relations between regional dimensions of poverty, social exclusion, and their predictors. Empirical analysis conducted at regional level can be potentially beneficial for the social policy making in any country when justifying a reform of – or a modern design for – the social system for poverty alleviation.

## REFERENCES

[1]. Boshnakov, V., Atanasov, A., Naydenov, A., and Chipeva, S. (2020). *Econometrics*. Sofia: UNWE Publishing.
[2]. De Haan, A., and Maxwell, S. (1998). *Poverty and Social Exclusion in North and Souht*. IDS Bulletin Vol.29 No.1.
[3]. Gordon, D., et al. (2000). *Poverty and Social Exclusion in Britain*. Joseph Rowntree Foundation.
[4]. Ismailov, T. (2019). Dimensions of Poverty and Social Exclusion among Bulgarian Population. *Science, Education, Society: Reality, Challenges, Perspectives*, Section 2, pp.76-79 (*jnos.donnu.edu.ua*).
[5]. MacInnes, T., Aldridge, H., Bushe, S., Kenway, P., and Tinson, A. (2013). *Monitoring Poverty and Social Exclusion 2013*. Joseph Rowntree Foundation.
[6]. Mintchev, V., Boshnakov, V., and Naydenov, A. (2010). Sources of Income Inequality: Empirical Evidence from Bulgaria. *Ikonomicheski Izsledvania*, 19 (4): 39-64.
[7]. Palmer, G., Rahman, M., and Kenway, P. (2002). *Monitoring Poverty and Social Exclusion 2002*. Joseph Rowntree Foundation
[8]. Stoyanova, R., Kokova, S., and Stoev, T. (2020). Risk of Poverty and Social Exclusion in Bulgaria: Socio-Demographic Determinants. *Management and Education*, 16 (1): 25-29.
[9]. Terziev, V., Arabska, E., Dimitrovski, R., Pushova, L. (2016). Challenges to social entrepreneurship development in Bulgaria. *11th International Scientific Conference Knowledge in Practice*, Bansko, Bulgaria. https://dx.doi.org/10.2139/ssrn.3155439.
[10]. Tinson, A., Ayrton, C., Barker, K., Born, T. B., Aldridge, H., and Kenway, P. (2016). *Monitoring Poverty and Social Exclusion 2016*. Joseph Rowntree Foundation.
[11]. United Nations (1995). *The Copenhagen declaration and programme of action: World summit for social development 6-12 March 1995*. New York: UN Department of Publications, p. 57.
[12]. Wooldridge, J. M. (2013). *Introductory Econometrics: A Modern Approach*. 5th Edn. Cengage Learning.
[13]. Zhelyazkova, M. (2018). *Policies towards Poverty in Bulgaria: Social Effectiveness and Necessary Reforms*. https://www.president.bg/docs/1540470457.pdf